\RequirePackage{fix-cm}
\documentclass[smallextended]{svjour3}       
\usepackage{natbib}

\smartqed  
\usepackage{graphicx}
\journalname{Empirical Software Engineering}
\usepackage{amsmath,amssymb,amsfonts}
\usepackage{algorithmic}
\usepackage{textcomp}
\usepackage{rotating}
\usepackage{multirow}
\usepackage[dvipsnames]{xcolor}

\usepackage[nolist]{acronym}
\usepackage{caption,booktabs}
\usepackage{hyperref}

\newcommand{\numberProjects}{23}
\newcommand{\numberIssuesLabeled}{1,723}  
\newcommand{\numberPMDWarnings}{314}  
\newcommand{\numberDefaultWarnings}{49}  

\begin{document}

\begin{acronym}
\acro{LLOC}{Logical Lines of Code}
\acro{ITS}{Issue Tracking System}
\acro{VCS}{Version Control System}
\acro{CI}{Continuous Integration}
\acro{AST}{Abstract Syntax Tree}
\acrodefplural{AST}[ASTs]{Abstract Syntax Trees}
\acro{DAG}{Directed Acyclic Graph}
\acro{ASAT}[ASAT]{Automated Static Analysis Tool}
\acrodefplural{ASAT}[ASATs]{Automated Static Analysis Tools}
\end{acronym}

\title{Are automated static analysis tools worth it?\\An investigation into relative warning density and external software quality}

\titlerunning{Are automated static analysis tools worth it?}        

\author{Alexander Trautsch \and
        Steffen Herbold \and
        Jens Grabowski
}

\institute{Alexander Trautsch\\Institute of Computer Science, University of Goettingen, Germany\\
           \email{alexander.trautsch@cs.uni-goettingen.de}
           \vspace{5pt}\\
           Steffen Herbold\\Institute of Software and Systems Engineering, TU Clausthal, Germany\\
           \email{steffen.herbold@tu-clausthal.de}
           \vspace{5pt}\\
           Jens Grabowski\\Institute of Computer Science, University of Goettingen, Germany\\
           \email{grabowski@cs.uni-goettingen.de}
}

\date{Received: date / Accepted: date}

\maketitle

\begin{abstract}
Automated Static Analysis Tools (ASATs) are part of software development best practices.
ASATs are able to warn developers about potential problems in the code.
On the one hand, ASATs are based on best practices so there should be a noticeable effect on software quality.
On the other hand, ASATs suffer from false positive warnings, which developers have to inspect and then ignore or mark as invalid.
In this article, we ask the question if ASATs have a measurable impact on external software quality, using the example of PMD for Java.
We investigate the relationship between ASAT warnings emitted by PMD on defects per change and per file.
Our case study includes data for the history of each file as well as the differences between changed files and the project in which they are contained.
We investigate whether files that induce a defect have more static analysis warnings than the rest of the project.
Moreover, we investigate the impact of two different sets of ASAT rules.
We find that, bug inducing files contain less static analysis warnings than other files of the project at that point in time. However, this can be explained by the overall decreasing warning density.
When compared with all other changes, we find a statistically significant difference in one metric for all rules and two metrics for a subset of rules.
However, the effect size is negligible in all cases, showing that the actual difference in warning density between bug inducing changes and other changes is small at best.
\end{abstract}

\section{Introduction}
\acp{ASAT} or linters are programs that perform rule matching of source code via different representations, e.g., \acp{AST}, call graphs or bytecode to find potential problems.
Rules are predefined by the \ac{ASAT} and based on common coding mistakes and best practices. If a rule is matched, a warning is generated for the developer who can then inspect the given file, line number and rule.
Common coding best practices involve \acp{ASAT} use at different contexts~\citep{Vassallo2020}, e.g., as part of \ac{CI}, within IDEs, or to support code reviews.
Developers also think of these tools as quality improving when used correctly~\citep{Christakis2016,Vassallo2020,Devanbu2016,Querel2021}.
However, due to their rule matching nature, \acp{ASAT} are prone to false positives, i.e., warnings about code that is not problematic~\citep{whynot_static}.
This hinders the adoption of these tools and their usefulness, as developers have to inspect every warning that is generated whether it is a false positive or not.
As a result, research into classifying \ac{ASAT} warnings into true and false positives or actionable warnings is conducted, e.g.,~\cite{Heckman2009}, \cite{warnings_static} and \cite{false_positives}.
Due to these two aspects, \acp{ASAT} are perceived as quality improving while at the same time require manual oversight and corrections.

Due to this manual effort, we want to have a closer look on the impact of \acp{ASAT} on measurable quality.
Previous research regarding the impact on quality can be divided into direct measures which target bug fixing commits, e.g.,~\cite{vetro2011}, \cite{thung,habib} and
indirect measures which use \ac{ASAT} warnings as part of predictive models, e.g.,~\cite{bugfinding_static}, \cite{static_workshop}, \cite{defect_prediction_findbugs}, \cite{Querel2018}, \cite{Lenarduzzi2020}, \cite{Trautsch2020a} and \cite{Querel2021}.
Both approaches usually suffer from data validation issues. \ac{ASAT} warnings measured directly in bug fixing commits are usually validated via manual inspection of the warnings either by students~\citep{vetro2011} or by the researchers themselves~\citep{thung,habib}.
Predictive models that include \ac{ASAT} warnings as features includes bugs and bug fixing commits which introduces new data validity problems, e.g., with noisy issue types as reported by~\cite{Antoniol2008} and \cite{herzigbug}, links from bug reports to commits, or the SZZ~\citep{szz} implementation~\citep{Fan2019}.
Recently \cite{Rosa2021} demonstrated in their investigation that finding bug inducing information is challenging, when using common or even state-of-the art SZZ approaches.  
In this article, we mitigate this by using manually validated bug fixing data from a large-scale validation study~\citep{line_label}.
While this reduces the data available for the study, we believe that this is a worthwhile trade off to be able to get more signal with less noise from the available data.

Between direct and indirect impact studies is a missing piece however.
Direct impact studies make sense for security focused \acp{ASAT} (Flawfinder, RATS) or bug focused \acp{ASAT} (FindBugs, SpotBugs), but not necessarily for general purpose \acp{ASAT} (PMD, SonarQube).
General purpose \acp{ASAT} are able to uncover readability related issues, e.g., style or naming issues in addition to common programming errors or violations of best practices.
These issues, while possibly not directly responsible for a bug, may introduce bugs later, or, in other parts of the file due to the reduced understandability of the code.
Indirect studies that rely on ASAT warnings as features for defect prediction may be too indirect to measure the actual relation between the \acp{ASAT} and defects,
as this is by necessity done in relation to prediction models using other features of changes. Moreover, this approach ignores problems due to differences in warning density between projects or correlations with size or time.

In order to make general statements, differencing techniques can be used, e.g., differences between bug inducing commits and previous commits as is done in the work by \cite{Lenarduzzi2020}.
What is missing however, is a more general view of the differences between warning densities of the files and the rest of the project at each time step.
Such a view would allow us to explore whether files that contain more static analysis warnings than the rest of the project also induce more bugs.

In prior work~\citep{Trautsch2020} we investigated trends of ASAT warnings and investigated whether usage of an ASAT has an influence on software quality measured via defect density~\citep{fenton}.
However, our analysis with regards to bugs via defect density was coarse grained.
We address this limitation in this article.
We adopt a recently introduced approach for fine-grained just-in-time defect prediction by~\cite{Pascarella2019} to perform a more targeted investigation of the impact of static analysis warnings on quality via introduced bugs as a proxy for software quality instead of the more coarse grained defect density for one year.
The approach by~\cite{Pascarella2019} provides the advantage of handling files within commits and not only commits which makes it suitable for us to study the impact of static analysis over the history of all files within our study subjects.

In our previous work, we found that static analysis warnings are evolving over time and that we can not just use the density or the sum of warnings in our case study~\citep{Trautsch2020}.
Therefore, we use an approach that is able to produce a current snapshot view of the files we are interested in by measuring the file that induces a bug and, at the same time, all other files of the project.
This ensures that we are able to produce time and project independent measurements.
The drawback of this approach is that it requires a large computational effort, as we have to run the \ac{ASAT} under study on every file in every revision of all study subjects.
However, the resulting empirical data yields insights for researchers and practitioners.\\

\noindent
The research question that we answer in our exploratory study is:
\begin{itemize}
    \item Do bug inducing files contain more static analysis warnings than other files?
\end{itemize}
We apply a modified fine-grained just-in-time defect prediction data collection method to extract software evolution data including bug inducing file changes and static analysis warnings from a general purpose ASAT.
We chose PMD\footnote{https://pmd.github.io/} as the general purpose ASAT as it has been available for a long time and provides a good mix of available rules.
Using this data and a warning density based metric calculation, we investigate the differences between bug inducing files and the rest of the studied system at the point in time when the bug is introduced.
In summary, this article contains the following contributions.
\begin{itemize}
    \item A unique and simple approach to measure impact of \acp{ASAT} that is independent of differences between projects, size and time.
    \item Complete static analysis data for PMD for \numberProjects{} open source projects for every file in every commit.
    \item An investigation into relative warning density differences within bug inducing changes.
\end{itemize}
The main findings of our exploratory study are:
\begin{itemize}
    \item Bug inducing files do not contain higher warning density than the rest of the project at the time when the bug is introduced.
    \item When comparing bug inducing warning density with all other changes we can measure higher warning density on a subset of PMD warnings that is a popular default for two metrics and for all available rules for one metric.
\end{itemize}

\noindent
The rest of this article is structured as follows.
Section~\ref{sec:related_work} lists previous research related to this article and discusses the differences.
Section~\ref{sec:case_study} describes the case study setup, methodology, analysis procedure and the results.
Section~\ref{sec:discussion} discusses the results of our case study and relates them to the literature.
Section~\ref{sec:threats_to_validity} lists and discusses threats to validity we identified for our study.
Section~\ref{sec:conclusion} concludes the article with a short summary and provides a short outlook.

\section{Related Work}\label{sec:related_work}
In this article, we explore a more general view of \acp{ASAT} and the warning density differences of bug inducing changes.
This can be seen as a mix of a direct and indirect impact study. Therefore we describe related work for both direct and indirect impact studies within this section.

The direct impact is often evaluated by exploring if bugs that are detected in a project are fixed by removing \ac{ASAT} warnings, i.e., did the warning really indicate a bug that needed to be fixed later.

\cite{thung} investigated bug fixes of three open source projects and three \acp{ASAT}: PMD, JLint, and FindBugs.
The authors look at how many defects are found fully and partially by changed lines and how many are missed by the \acp{ASAT}.
Moreover, the authors describe the challenges of this approach: not every line that is changed is really a fix for the bug, therefore the authors perform manual investigation on a per-line level to identify the lines.
They were able to find all lines responsible for 200 of 439 bugs. 
In addition, the authors find that PMD and FindBugs perform best, however their warnings are often very generic.

\cite{habib} perform an investigation of the capabilities to find real world bugs via \acp{ASAT}. The authors used the Defects4J dataset by~\cite{defects4j} with an extension\footnote{https://github.com/rjust/defects4j/pull/112} to investigate the number of bugs found by three static analysis tools, SpotBugs, Infer and error-prone.
The authors show that 27 of 594 bugs are found by at least one of the \acp{ASAT}.

In contrast to \cite{thung} and \cite{habib}, we only perform an investigation of PMD. However, due to our usage of SmartSHARK \citep{smartshark}, we are able to investigate \numberIssuesLabeled{} bugs for which at least three researchers achieved consensus
on the lines responsible for the bug. Moreover, as PMD includes many rules related to readability and maintainability, we build on the assumption that while they are not directly indicating a bug, resolving these warnings improves the quality of the code and may prevent future bugs. This extends previous work by taking possible long term effects of \ac{ASAT} warnings into account.

Indirect impact is explored by using \ac{ASAT} warnings as features for predictive models and providing a correlation measure of \ac{ASAT} warnings to bugs.

\cite{bugfinding_static} explore the ability of \ac{ASAT} warnings to predict defect density in modules.
The authors found in a case study with Microsoft, that static analysis warnings can be used to predict defect density, therefore they can be used to focus quality assurance efforts on modules that
show a high number of static analysis warnings.
In contrast to \cite{bugfinding_static}, we are exploring open source projects.
Moreover, we explore warning density differences between files and the project they are contained in.

\cite{defect_prediction_findbugs}~compare static analysis and statistical defect prediction.
They find that FindBugs is able to outperform statistical defect prediction, while PMD does not.
Within our study, we focus on PMD as a general purpose \ac{ASAT}. Instead of a comparison with statistical defect prediction we explore, whether we can measure a difference of \ac{ASAT} warnings between bug inducing changes and other changes.

\cite{static_workshop} explores a correlation between \ac{ASAT} warnings as features for a predictive model and the dependent variable, i.e., bugs.
They found that static analysis warnings may improve the performance of predictive models and that they are correlated with bugs.
In contrast to~\cite{static_workshop}, we are not building a predictive model.
We are exploring whether we can find an effect of static analysis tools without a predictive model with multiple features, instead we strive to keep the approach as simple as possible.

\cite{Querel2018} improve the just-in-time defect prediction based commit guru~\citep{rosen2015} by adding \ac{ASAT} warnings to the predictive model.
The authors show, that just-in-time defect prediction can be improved by adding static analysis warnings.
This means that there should be a connection between external quality in the form of bugs and static analysis warnings.
In a follow up study~\citep{Querel2021} the authors found that while there is an effect of \ac{ASAT} warnings the effect is likely small.
In our study, we explore a different view on the data. We explore warning density differences between bug inducing files and the rest of the project.

\cite{Lenarduzzi2020}~investigated SonarQube as an ASAT and if the reported warnings can be used as features to detect reported bugs.
The authors are combining direct with indirect impact but are more focused on predictive model performance measures.
In contrast to \cite{Lenarduzzi2020}, we are mainly interested in the differences in warning density between bug inducing files and the rest of the project.
We are also investigating an influence, but in contrast to Lenarduzzi et al., we are comparing our results for bug fixing changes to all other changes to determine whether what we see is really part of the bug fixing change and not a general trend of all changes.

\section{Case Study}\label{sec:case_study}
The goal of the case study is to find evidence if usage of \acp{ASAT} have a positive impact on the external software quality of our case study subjects.
In this section, we explain the approach and \ac{ASAT} choice.
Moreover, we explain our study subject selection and describe the methodology and analysis procedure.
At the end of this section we present the results.

\subsection{Static analysis}
Static analysis is a programming best practice.
\acp{ASAT} scan source code or byte code and match against a predefined set of rules. When a rule matches, the tool creates a warning for the part of the code that matches the rule.

There are different tools for performing static analysis of source code. For Java these would be, e.g., Checkstyle, FindBugs/SpotBugs, PMD, or SonarQube.
In this article, we focus on Java as a programming language because it is widely used in different domains and has been in use for a long time.
The static analysis tool we use is PMD. There are multiple reasons for this.
PMD does not require the code to be compiled first as, e.g., FindBugs does. This is an advantage especially with older code that might not compile anymore due to missing dependencies~\citep{Tufano2017}.
PMD supports a wide range of warnings of different categories, e.g., naming and brace rules as well as common coding mistakes. This is an advantage over, e.g., Checkstyle which mostly deals with coding style related rules.
This enables PMD to give a better overview of the quality of a given file instead of giving only probable bugs within it.
The relation to software quality that we expect of PMD stems directly from its rules. The rules are designed to make the code more readable, less error prone and overall more maintainable.

\subsection{Just-in-time defect prediction}
The idea behind just-in-time defect prediction is to assess the risk of a change to an existing software project~\citep{Kamei2013}.
Previous changes are extracted from the version control system of the project and, as they are in the past, it is known whether the change induced a bug.
This can be observed by subsequent removal or alteration of the change as part of a bug fixing commit.
If the change was indeed removed or altered as part of a bug fixing operation it is traced back to its previous file and change and labeled as bug inducing, i.e., it introduced a bug that needed to be fixed later.
In addition to these labels, certain characteristics of the change are extracted as features, e.g., lines added or the experience of the author to later train a model to predict the labels correctly for the commits.
The result of the model is then a label or probability whether the change introduces a bug, i.e., the risk of the change.

However, \acp{ASAT} are working on a file basis and we also want to investigate longer-term effects of \acp{ASAT}. This means we need to track a file over its evolution in a software project.
To achieve this, we are building on previous work by~\cite{Pascarella2019} which introduced fine-grained just-in-time defect prediction.
In a previous study, we improved the concept by including better labels and static analysis warnings as well as static code metrics as features~\citep{Trautsch2020a}.
Similar to \cite{Pascarella2019}, we are building upon PyDriller \citep{pydriller}.
In this article, we build upon our previous work and include not only counts of static analysis warnings but relations between the files, e.g., how different is the number of static analysis warnings in one file from the rest of the project.
We also include aggregations of warnings with and without a decay over time.

\subsection{Study Subjects}
Our study subjects consist of \numberProjects{} Java projects under the umbrella of the Apache Software Foundation\footnote{https://www.apache.org} previously collected by~\citep{Datensatz}.
Table~\ref{tbl:study_subjects} contains the list of our study subjects.
We only use projects which contain fully validated bug fixing on a line-by-line level collected in a crowd sourcing study~\citep{line_label}.
Every line in our data was labeled by four researchers. We only consider bug fixing lines for which at least three researchers agree that it fixes the considered bug.
This naturally restricts the number of available projects but improves the noise to signal ratio of the data.
\begin{table}
    \centering
    \caption{Study subjects in our case study}\label{tbl:study_subjects}
    \begin{tabular}{lrrrr}
        \toprule
        Project & \#commits & \#file changes & \#issues & Time frame\\
        \midrule
ant-ivy & 1,647 & 7,860 & 296 & 2005-2017\\
commons-bcel & 850 & 9,604 & 27 & 2001-2017\\
commons-beanutils & 561 & 2,648 & 28 & 2001-2017\\
commons-codec & 810 & 2,062 & 21 & 2003-2017\\
commons-collections & 1,687 & 11,296 & 32 & 2001-2017\\
commons-compress & 1,401 & 3,566 & 87 & 2003-2017\\
commons-configuration & 1,659 & 4,177 & 97 & 2003-2017\\
commons-dbcp & 729 & 2,211 & 39 & 2001-2017\\
commons-digester & 1,131 & 3,750 & 11 & 2001-2017\\
commons-io & 985 & 2,781 & 51 & 2002-2017\\
commons-jcs & 774 & 7,775 & 37 & 2002-2017\\
commons-lang & 3,028 & 6,312 & 109 & 2002-2017\\
commons-math & 4,135 & 21,440 & 190 & 2003-2017\\
commons-net & 1,076 & 4,666 & 96 & 2002-2017\\
commons-scxml & 469 & 1,774 & 39 & 2005-2017\\
commons-validator & 557 & 1,324 & 37 & 2002-2017\\
commons-vfs & 1,098 & 7,209 & 67 & 2002-2017\\
giraph & 819 & 7,715 & 109 & 2010-2017\\
gora & 464 & 2,256 & 38 & 2010-2017\\
opennlp & 1,166 & 9,679 & 82 & 2010-2017\\
parquet-mr & 1,053 & 5,957 & 69 & 2012-2017\\
santuario-java & 1,177 & 8,503 & 41 & 2001-2017\\
wss4j & 1,711 & 12,218 & 120 & 2004-2017\\
        \midrule
Sum & 28,987 & 146,783 & 1,723 & \\
        \bottomrule
    \end{tabular}
\end{table}
We now give a short overview what the potential problems are and how we mitigate them.
When we look at external quality, we want to extract data about defects. However, there are several additional restrictions we want to apply.
First, we want to extract defects from the \ac{ITS} of the project and link them to commits in the \ac{VCS} to determine bug fixing changes. Several data validity considerations need to be taken in to account here.
The \ac{ITS} usually has a kind of label or type to distinguish bugs from other issues, e.g., feature requests. However, research shows that this label is often incorrect, e.g.,~\cite{Antoniol2008}, \cite{Herzig2013} and \cite{Datensatz}.
Moreover, with this kind of software evolution research, we are interested in bugs existing in the software and not bugs which occur because of external factors, e.g., new environments or dependency upgrades.
Therefore, we are only considering intrinsic bugs~\citep{Extrinsic}.

The next step is the linking between the issue from the \ac{ITS} and the commit from the \ac{VCS}. This is achieved via a mention of the issue in the commit message, e.g., fixes JIRA-123.
While this seems straightforward there are certain cases where this can be problematic.
The simplest one being that there is a typo in the project key, e.g., JRIA-123.

Moreover, not all changes within bug fixing commits contribute to bug fixes. Unrelated changes can be tangled with the bug fix.
The restriction of all data to only changes that directly contribute to the bug fix further reduces noise in the data.
We are only interested in the lines of the changes that contribute to the bug fix. This is probably the hardest to manually validate.

This was achieved in a prior publication~\citep{Datensatz} which served as the base for the publication which data we use in this article~\citep{line_label}.
In~\citep{line_label} a detailed untangling is performed by four different persons for each change that fixes a bug that meets our criteria.

\subsection{Replication Kit}
We provide all data and scripts as part of a replication kit\footnote{https://github.com/atrautsch/emse2021a\_replication}.

\subsection{Methodology}
To answer our research question, we extract information about the history of our study subjects including bugs and the evolution of static analysis warnings.
While the bulk of the data is based on~\citep{line_label} we include several additions necessary for answering our research question.

To maximize the relevant information within our data we include as much information from the project source code repository as possible.
After extracting the bug inducing changes, we build a commit graph of all commits of the project and then find the current main branch, usually master.
After that, we find all orphan commits, i.e., all commits without parents. Then we discard all orphans that do not have a path to the last commit on the main branch, this discards separate paths in the graph, e.g., gh-pages\footnote{https://docs.github.com/en/pages} for documentation.
As we also want to capture data on release branches which are never merged back into the main branch, we add all other branches that have a path to one of our left over orphan commits.
The end result is a connected graph which we traverse via a modified breadth first search. We take the date of the commit into account while we traverse the graph.

The traversal is an improved version of previous work~\citep{Trautsch2020a}.
In addition to the previously described noise reduction via manual labeling, we additionally restrict all files to production code.
One of the results of~\cite{line_label} is that non-production code is often tangled with bug fixing changes.
Therefore we only add files that are production files to our final data analogous to~\cite{Trautsch2020}.
This also helps us to provide a clearer picture of warning density based features as production code may have a different evolution of warning density than, e.g., test or example code.

In our previous study~\citep{Trautsch2020} we found that static analysis warnings are correlated to \ac{LLOC}. This is not surprising as we are observing large portions of our study subjects code history.
Large files that are added and removed have an impact on the number of static analysis warnings. While we do not want to discard this information
we 
also want to avoid the problem of large changes overshadowing information in our data.
Therefore, like in our previous study we are using warning density as a base metric in this study analogous to prior studies, e.g., \cite{Aloraini2019} and \cite{Penta2009}. 

Warning density ($wd$) is the ratio of the number of warnings and the size of the analyzed part of the code.
\begin{equation}
	wd = \frac{\text{Number of static analysis warnings}}{\text{Product size}}
\end{equation}
Product size is measured in \ac{LLOC}. If we measure the warning density of a system $wd(s)$, we sum warnings and \ac{LLOC} for each file.
If we measure the warning density of a file $wd(f)$, we restrict the number of warnings and the \ac{LLOC} to that file.

While this measure provides a size independent metric, we also need to take differences between projects into account.
Warning density can be different between projects and even more so for different points in time for each project.
To be able to use all available data we account for these differences by using differences in warning density between the files of interest and the rest of the project under study (the system)
at the specific point in time.

We calculate the warning density difference between the file and the system~$fd(f_t)$.
\begin{equation}
    fd(f_t) = wd(f_t) - wd(s_t)
\end{equation}
If the file $f$ at time $t$ contains less static analysis warnings per LLOC than the system $s$ at time $t$ the value is negative and if it contains more it is positive.
We can use this metric to investigate bug inducing commits and determine whether the files responsible for bugs contain less or more static analysis warnings per LLOC than the system they belong to.

While this yields information corrected for size, project differences, and time of the change we also want to incorporate the history of each file.
Therefore, we also sum this difference in warning density for all changes to the file.
We assume that recent changes are more important than old changes. Therefore, we introduce a decay in our warning density derived features.
\begin{equation}
	dfd(f_t) = \sum_{j=1}^{j=t}{\frac{wd(f_j) - wd(s_j)}{t-j+1}}
\end{equation}
For the decayed file system warning density delta $dfd(f_t)$ we compute the decayed, cumulative sum of the difference between the warning density of the file ($wd(f_t)$) and the warning density of the system ($wd(s_t)$).
The rationale is that if a file is constantly better, with regards to static analysis warnings, than the mean of the rest of the system this should have a positive effect. As the static analysis rules are diverse this can be improved readability, maintainability or robustness due to additional null checks.
Within our study, we explore if this effect has a measurable effect on buggyness, i.e., the lower this value is the less often the file should be part of bug inducing commits.

Instead of using all warnings for warning density we can also restrict these warnings to a smaller set to see if this has an effect.
While we do not want to choose multiple subsets to avoid false positive findings, we have to investigate whether our approach to use all available warnings just waters down the ability to indicate files which may contain bugs.
To this end, we also investigate the warning density consisting only of PMD warnings that are enabled by default by the maven-pmd plugin\footnote{https://maven.apache.org/plugins/maven-pmd-plugin/} which we denote as \textit{default} rules.
This restricts the number of warnings that are the basis of the warning density calculation to a subset of \numberDefaultWarnings{} warnings that are generally considered relevant
in comparison to the total number of \numberPMDWarnings{} warnings.
Their use as default warnings serves to restrict this subset to generally accepted important warnings.

To answer our research question we compare the warning density for each bug inducing file against the project at the time before and after the bug inducing change.
If the difference is positive this means that the file had a higher warning density than the rest of the project and negative vice versa.
We plot the difference in warning density in a box plot for all bug inducing files to provide an overview over all our data.

As this is influenced by a continuously improving warning density we also measure the differences between bug inducing file changes and all other file changes.
We first perform a normality test and find that the data is not normal in all cases.
Thus, we apply a Mann-Whitney U test~\citep{mwu} with $H_0$ that there is no difference between both populations and $H_1$ that bug inducing files have a different warning density. We set a significance level of 0.05.
Additionally, we perform a Bonferroni~\citep{bonferroni} correction for 8 normality tests and 4 Mann-Whitney U tests. Therefore we reject $H_0$ at p $<$ 0.0042.
If the difference is statistically significant we calculate the effect size with Cliff's $\delta$~\citep{cliffsd}.

\subsection{Results}
We now present the results of our study and the answer to our
research question whether bug inducing files contain more static analysis warnings than other files.
For this, we divide the results into three parts. First, we look at the warning density via $fd(f)$ at the time before and after a bug is induced and $dfd(f)$ after a bug is induced\footnote{before is already part of the formula}.
Second, we look at the differences between our study subjects and the prior number of changes for bug inducing file changes.
Third, we compare bug inducing file changes with all other changes and determine if they are different.

\subsubsection{Differences of warning density before and after the bug inducing change}\label{sec:change}
\begin{figure}
    \centering
    \includegraphics[width=0.5\textwidth]{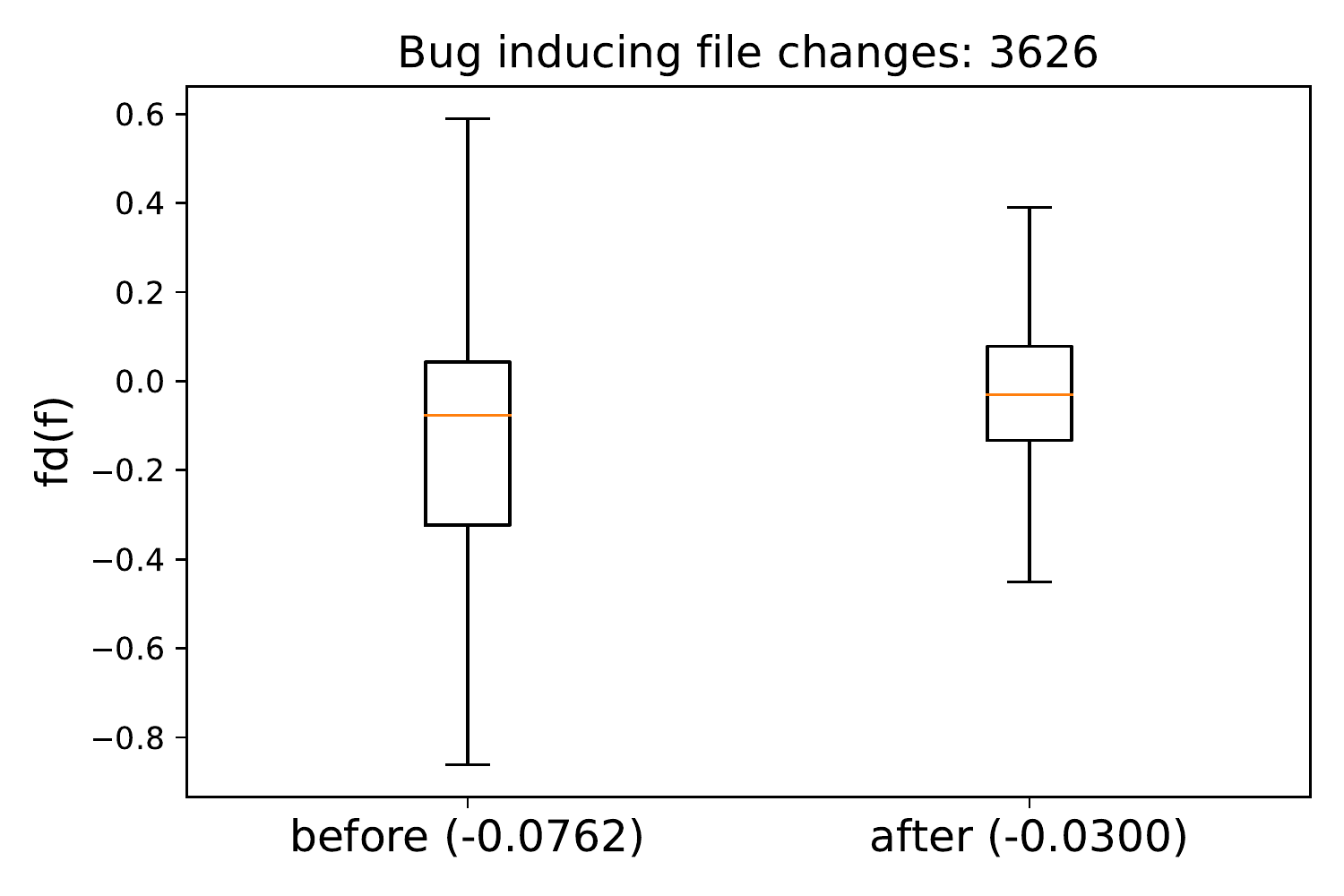}\includegraphics[width=0.5\textwidth]{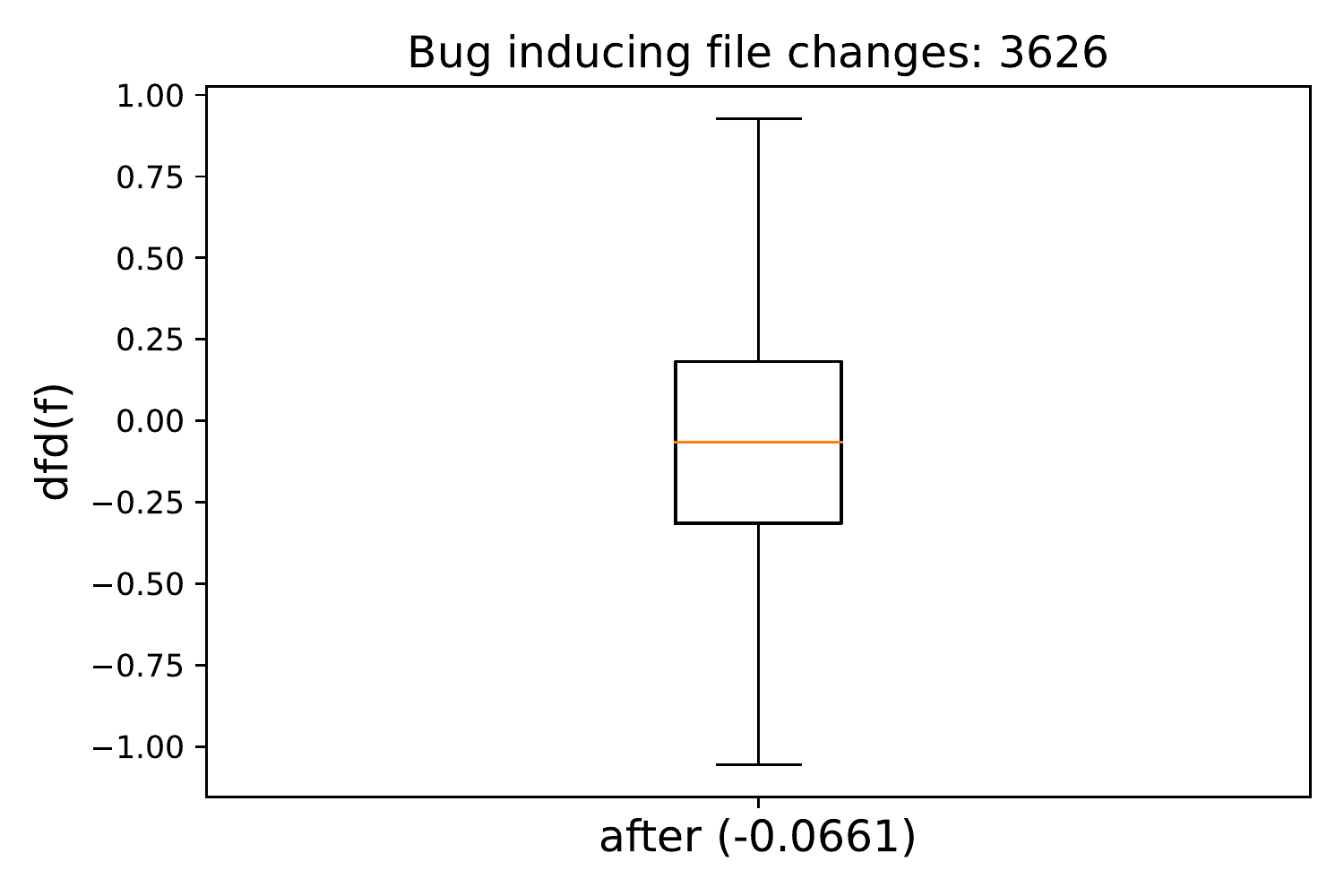}
    \caption{Box plot of $fd(f)$ for all bug inducing files before and after the bug inducing change and $dfd(f)$ for all bug inducing files after the bug inducing change, median value in parentheses. Fliers are omitted.}\label{fig:bp_rq1}
\end{figure}

Figure~\ref{fig:bp_rq1} shows the difference in warning density between the each bug inducing file and the rest of the system at the point in time before inducing the bug and after.
\begin{figure}
    \centering
    \includegraphics[width=0.5\textwidth]{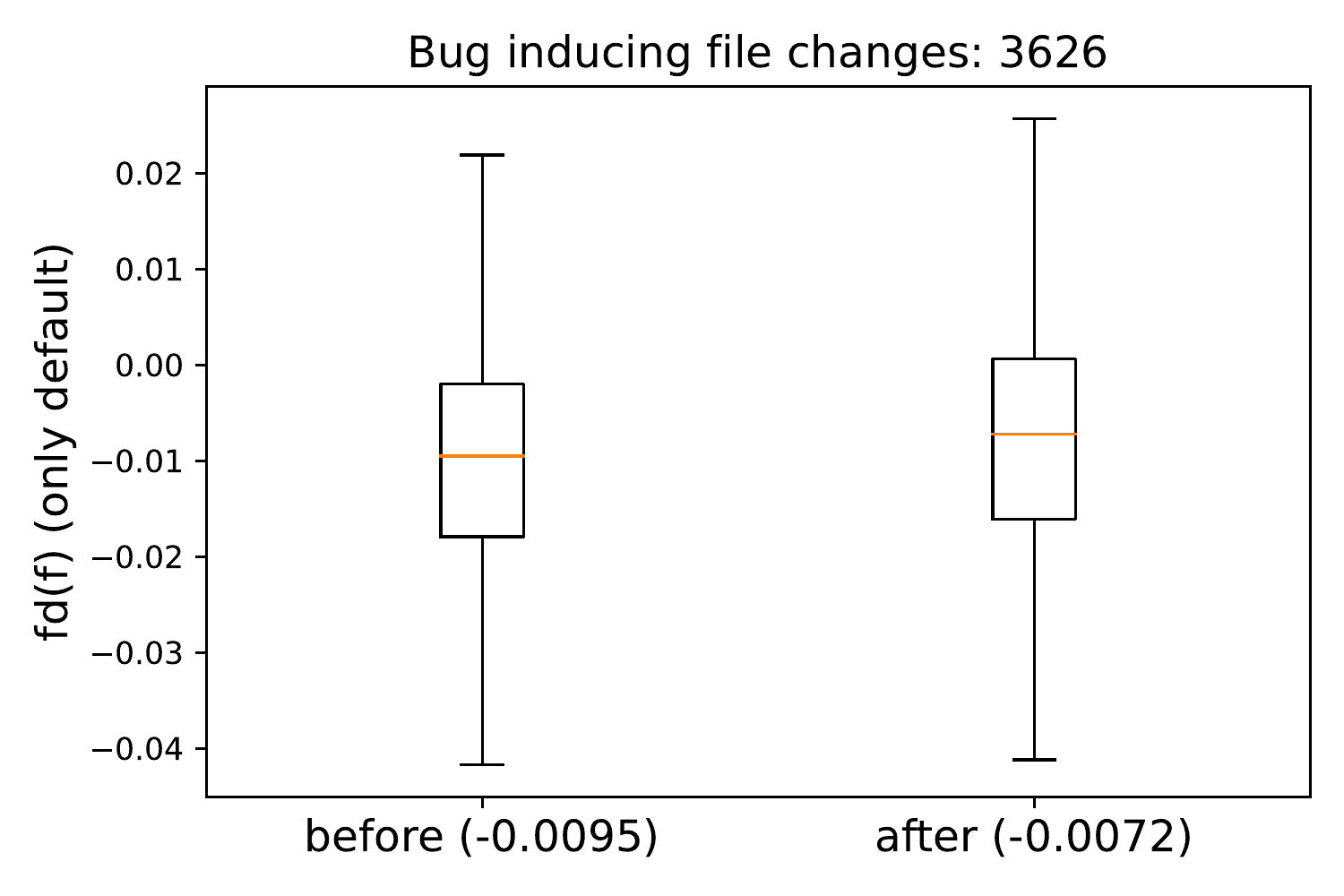}\includegraphics[width=0.5\textwidth]{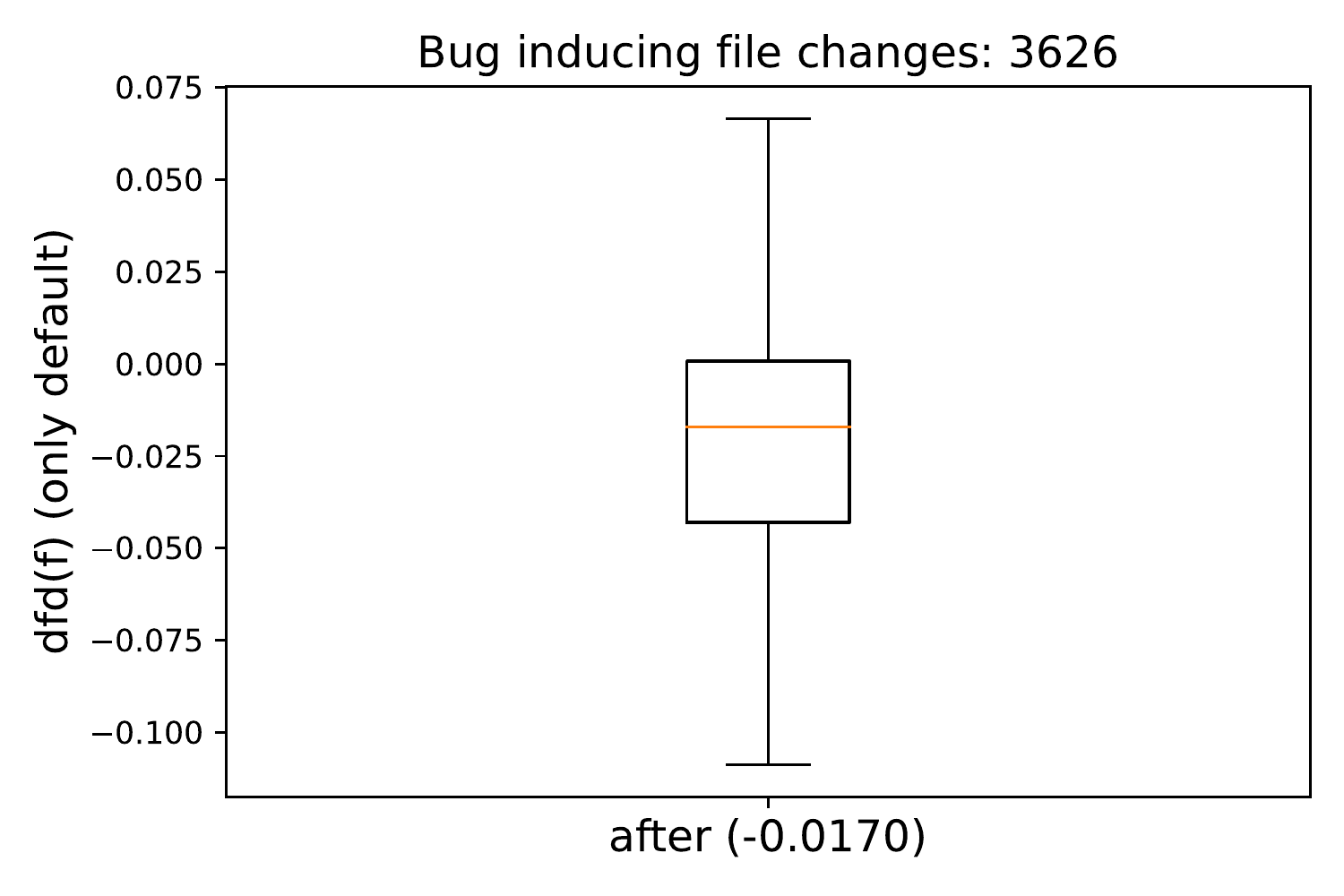}
    \caption{Box plot of $fd(f)$ for only default warnings of all bug inducing files before and after the bug inducing change, median value in parentheses. Fliers are omitted.}\label{fig:bp3_rq1}
\end{figure}
Surprisingly, we see a negative warning density median difference for $fd(f)$. This means that the warning density of the files in which bugs are induced is lower than the rest of the project.
The drop in warning density shows that the code before the change had less warning density than after the bug inducing change.
This means that code that on average contains more static analysis warnings was introduced as part of the bug inducing change.

Now, we are also interested if the history of preceding differences in warning density makes a difference.
Instead of using the warning density difference at the point in time of the bug inducing change we use a decayed sum of the warning density differences leading up to the considered bug inducing change.

Figure~\ref{fig:bp_rq1} shows a negative median for $dfd(f)$ as well. The accumulated warning density differences between the file and the rest of the project are therefore also negative.
Figure~\ref{fig:bp3_rq1} shows $fd(f)$ and $dfd(f)$ for bug inducing changes restricted to default rules. We can see, that the warning density for default only is much lower due to the lower number of warnings
that are considered. We can also see, that the same negative median is visible when we restrict the set of ASAT rules to default.
Overall, bug inducing changes have lower warning density than the other files of the project at the time the bug was induced.
However, as we will see later, this is an effect of overall decreasing warning density of our study subjects.

\subsubsection{Differences between projects and number of changes}\label{sec:diff_projects}
\begin{figure}
    \centering
    \includegraphics[width=0.9\textwidth]{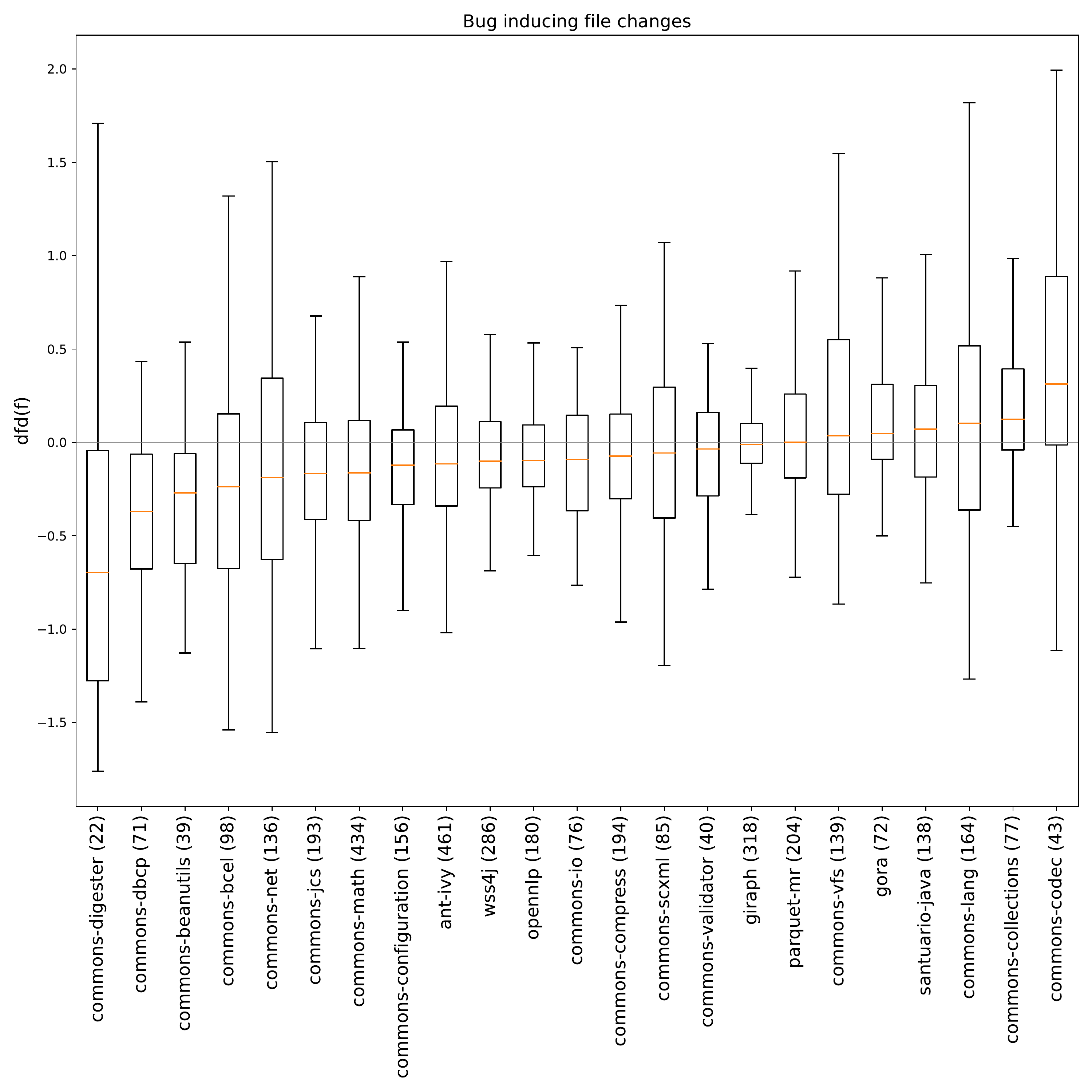}
    \caption{Box plots of $dfd(f)$ separately for all study subjects. The number of bug inducing file changes are in parentheses, median value in parentheses. Fliers are omitted.}\label{fig:all_projects}
\end{figure}
Instead of looking at all files combined we can also look at each project on its own. We provide this data in Figure~\ref{fig:all_projects}.
However, we note that the number of bug inducing files is low in some projects.
Such projects may be influenced by few changes with extreme values. Hence, the results of single projects should be interpreted with caution.
Instead we consider trends visible in the data.
While we can combine all our data due to our chosen method of metric calculation we still want to provide an overview of the per project values.
This is shown in Figure~\ref{fig:all_projects} for $dfd(f)$.
Figure~\ref{fig:all_projects} also demonstrates the difference between projects.
For example, the median $dfd(f)$ for comons-codec is is positive, i.e., files which induce bugs contain more warnings. The opposite is the case for, e.g., commons-digester, where the median is negative.

\begin{figure}
    \centering
    \includegraphics[width=0.5\textwidth]{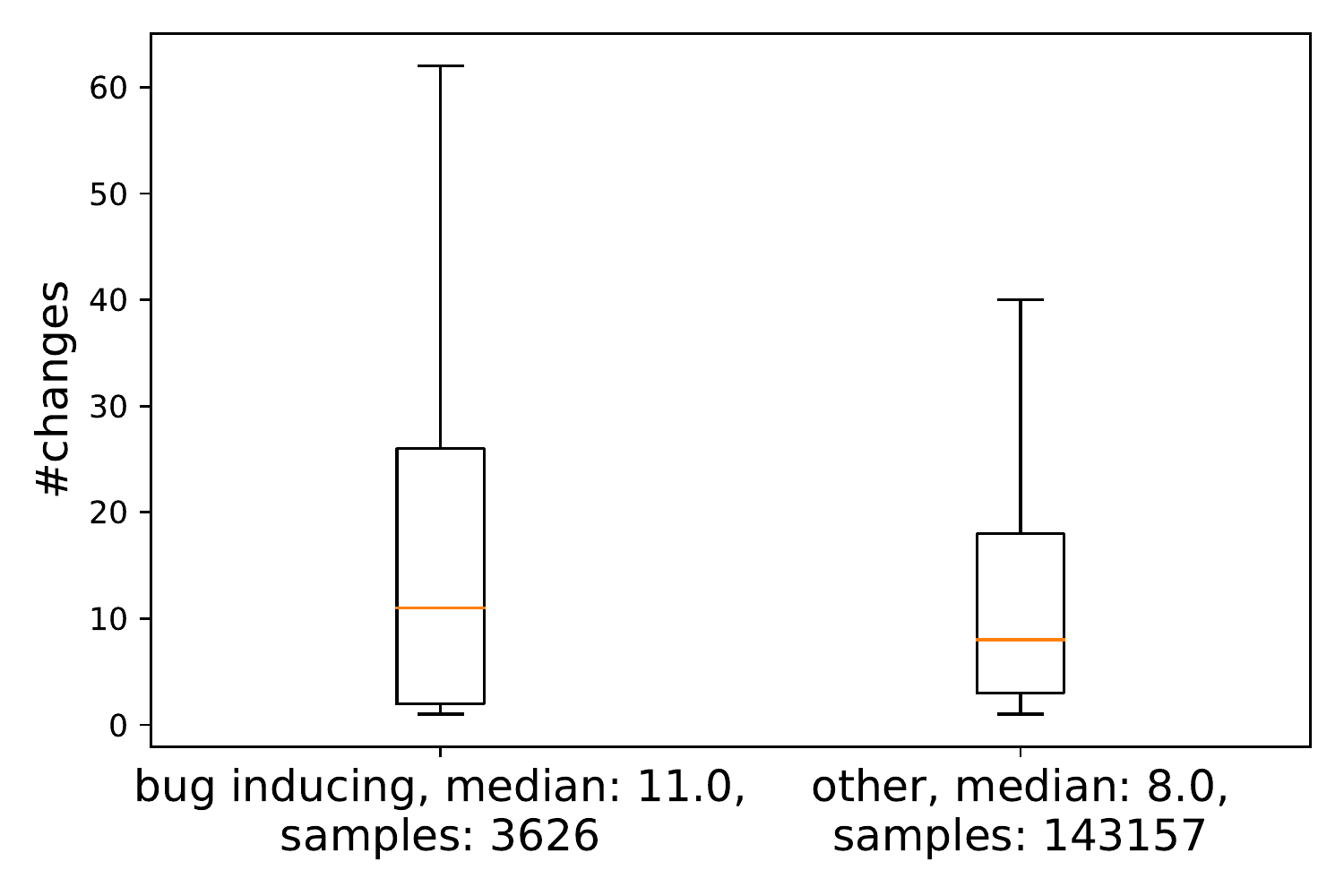}
    \caption{Number of changes for bug inducing files and other files. Fliers are omitted.}\label{fig:comm}
\end{figure}

Overall, Figure~\ref{fig:all_projects} shows that the median $dfd(f)$ is negative for 16 of 23 projects. This means that bug inducing changes have less warning density than the rest of the project for most study subjects.
A possible explanation for this could be that files which have a lower warning density are changed more often and those are the same that could be inducing bugs.
If we look at the number of changes a file has in Figure~\ref{fig:comm}, we can see that bug inducing files have a bit more changes.
However, the sample sizes for both are vastly different.

\subsubsection{Comparison with all other changes}\label{sec:comparison}
We now take a look at how warning density metrics differ in bug inducing changes from all other changes.
We notice that the median is below zero in all cases.
This is due to the effect that warning density usually decreases over time~\citep{Trautsch2020}.
Therefore, we provide a comparison of bug inducing changes with all other changes.

\begin{figure}
    \centering
    \includegraphics[width=0.45\textwidth]{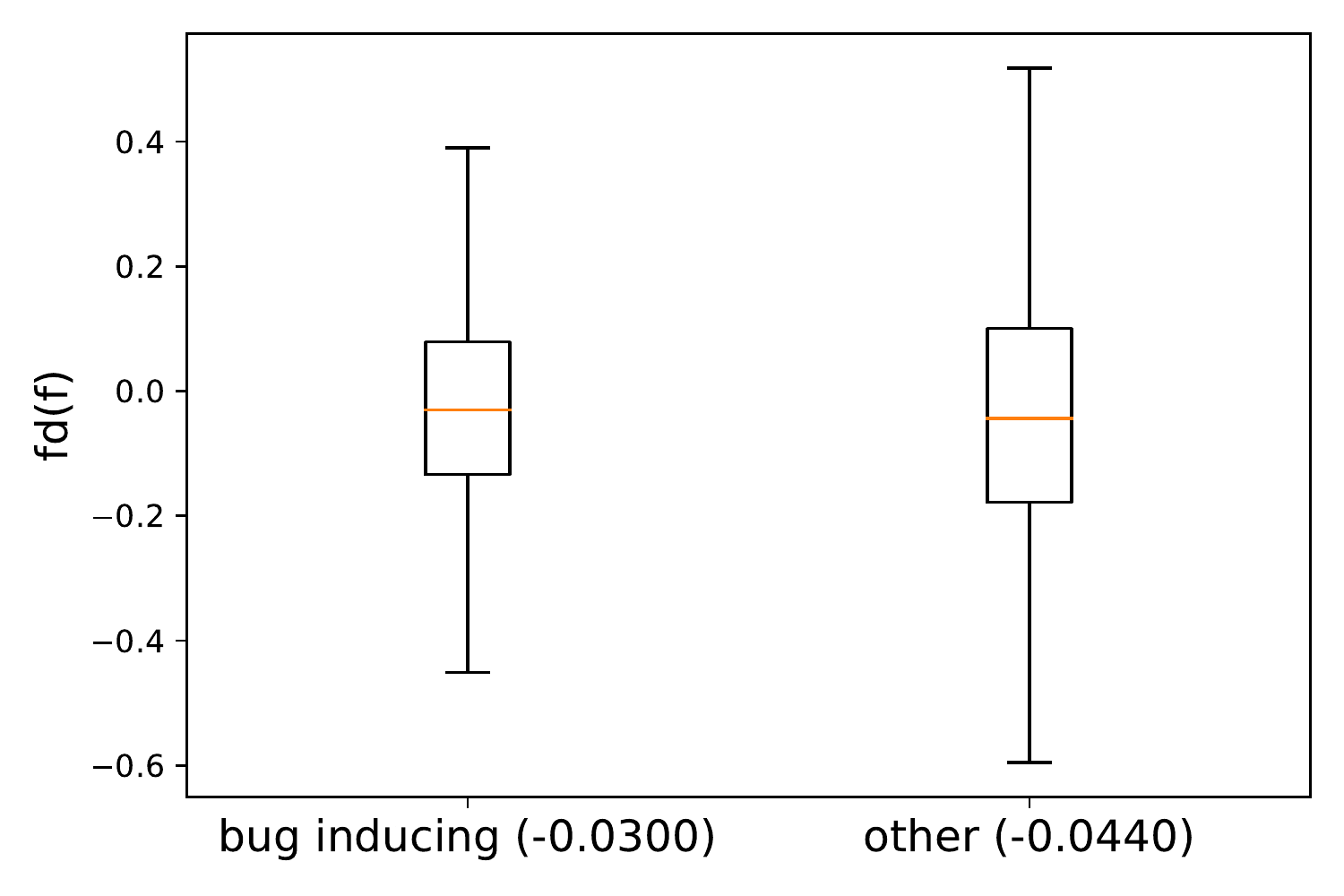}
    \includegraphics[width=0.45\textwidth]{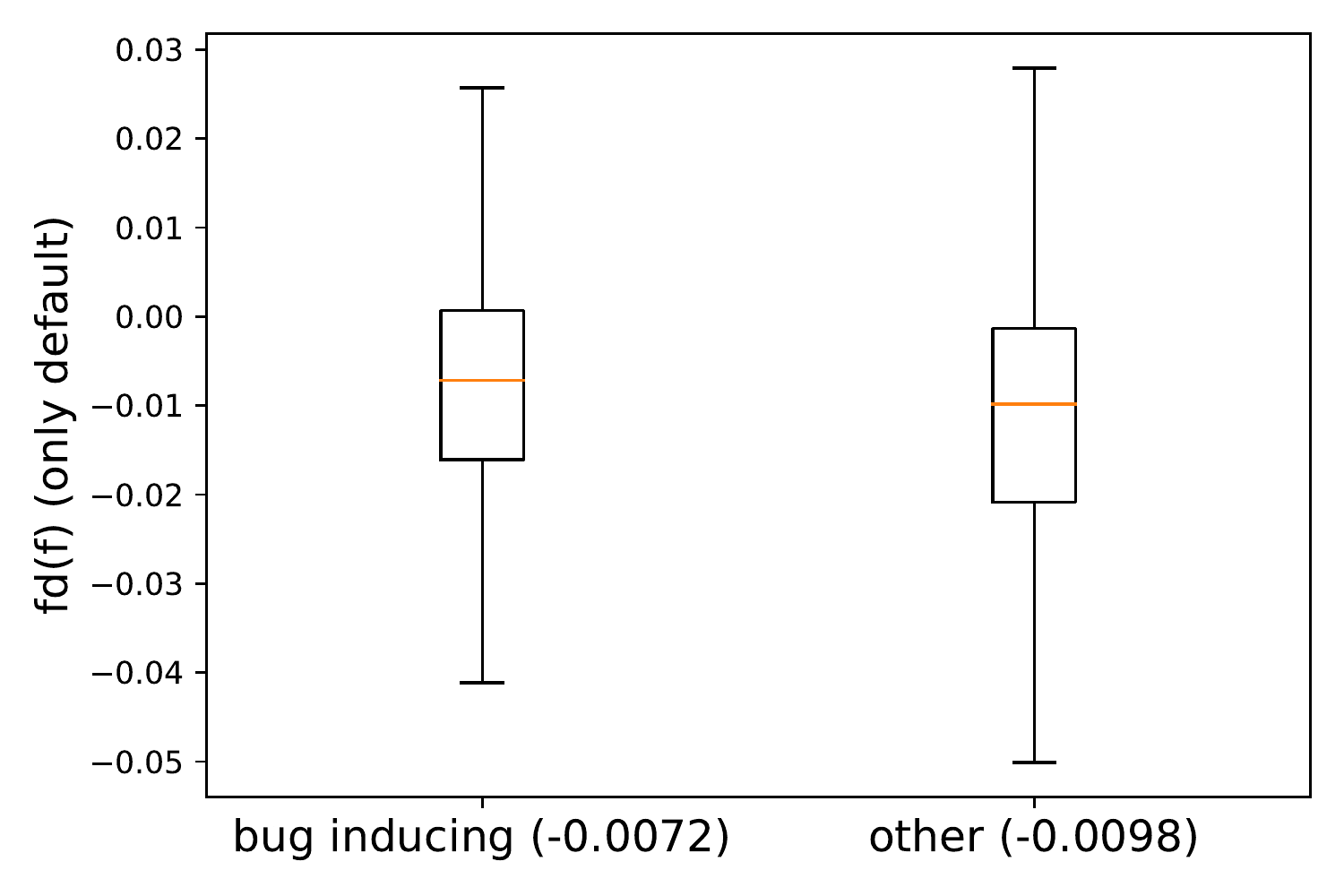}
    \caption{Box plots of $fd(f)$ before the bug inducing change for all and default only rules for bug inducing and other file changes, median value in parentheses. Fliers are omitted.}\label{fig:compare_wdf}
\end{figure}

Figure~\ref{fig:compare_wdf} shows $fd(f)$ for bug inducing and other changes for both all rules and only the default rules.
We can see that bug inducing changes have a slightly higher warning density than other changes. If we apply only default rules we see that bug inducing changes are also slightly higher.

\begin{figure}
    \centering
    \includegraphics[width=0.45\textwidth]{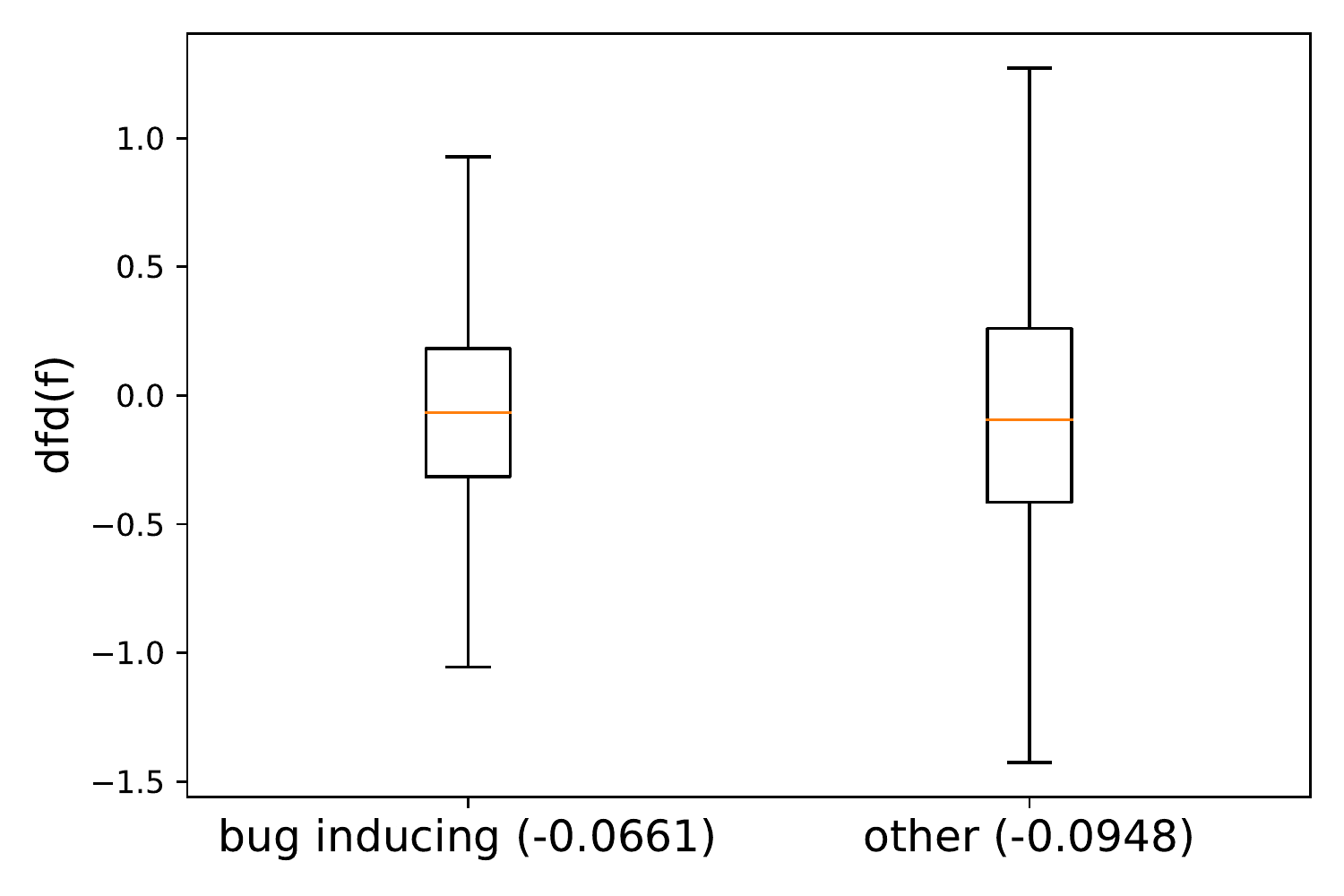}
    \includegraphics[width=0.45\textwidth]{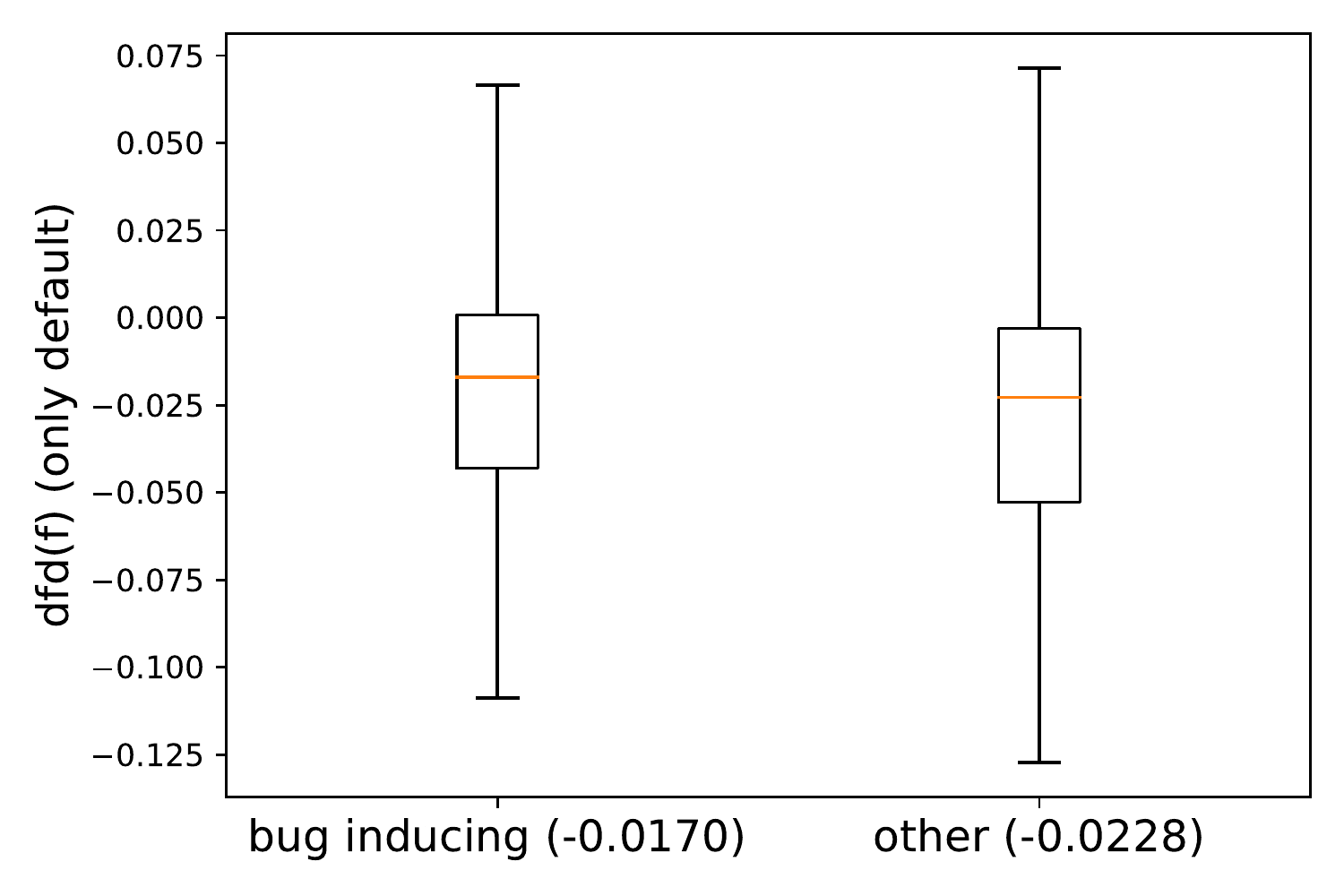}
    \caption{Box plots of $dfd(f)$ for all and default only rules bug inducing and other file changes, median value in parentheses. Fliers are omitted.}\label{fig:compare_dfd}
\end{figure}

Figure~\ref{fig:compare_dfd} shows the same comparison for $dfd(f)$. The difference for all rules is very small. However, the median for bug inducing changes is slightly higher. In contrast, we can see that for default rules the bug inducing changes have a slightly higher warning density than other changes.
Table~\ref{tbl:stats} shows the results of the statistical tests for differences between the values for Figure~\ref{fig:compare_wdf} and Figure~\ref{fig:compare_dfd}.

We can see that for all rules $fd(f)$ there is a statistically significant difference. This shows that bug inducing file changes have a higher warning density than other changes.

Overall, we see that there is a significant difference with a negligible effect size for $fd(f)$ and $dfd(f)$ for default rules between bug inducing and other changes.
The data shows that in these cases the bug inducing file changes have a higher warning density than other changes.
\begin{table}
    \centering
    \caption{Median values, Mann-Whitney U test p-values and effect sizes for all warning density metrics.}\label{tbl:stats}
    \begin{tabular}{lrrrr}
        \toprule
        WD Metric & Median other & Median bug inducing & P-value & Effect size\\
        \midrule
         $fd(f)$ & -0.0440 & -0.0300 & \textbf{\textless0.0001} & 0.05 (n)\\
         $fd(f)$ (default) & -0.0098 & -0.0072 & \textbf{\textless0.0001} & 0.10 (n)\\
         $dfd(f)$ & -0.0948 & -0.0661 & 0.0247 & -\\
         $dfd(f)$ (default) & -0.0228 & -0.0170 & \textbf{\textless0.0001} & 0.07 (n)\\
        \bottomrule
    \end{tabular}
\end{table}
Together with Figure~\ref{fig:compare_wdf}, Figure~\ref{fig:compare_dfd} and Table~\ref{tbl:stats} we can conclude, that bug inducing file changes contain more static analysis warnings than other file changes.
Restricting the rules to the default set increases the effect size slightly.
However, the effect sizes are still negligible in all cases.

\subsubsection{Summary}
In summary, we have the following results for our research question.
\begin{center}
    \setlength{\fboxrule}{0pt}
        \vrule\vrule\vrule\vrule\vrule\vrule\fcolorbox{white}{gray!15}{
            \hspace{.2em}\parbox{.91\linewidth} {
                \vspace{.5em}
                \textbf{RQ Summary}: Do bug inducing files contain more static analysis warnings than other files?\\

We find that bug inducing files contain less static analysis warnings than other files at bug inducing time.
However, this is not because these files have a higher quality, but rather because the warning density decreases over time, i.e., most files that are changed have less warnings than the rest of the project.\\

When we compare the differences between bug inducing and other changes, we find that it depends on the applied rules.
If we apply all rules we find a statistically significant difference in $fd(f)$.
If we apply only default rules we find a statistically significant difference in $fd(f)$ and $dfd(f)$.
This indicates that bug inducing files have slightly more static analysis warnings than other files, although the effect size in all cases is negligible.
                \vspace{.5em}
        }\hspace{.5em}
    }
\end{center}

\section{Discussion}\label{sec:discussion}
We found that the bug inducing change itself increased the warning density of the code in comparison to the rest of the project as shown in Figure~\ref{fig:bp_rq1}.
This means that the actual change in warning density is as we expected, i.e., the change that induces the bug is increasing the warning density in comparison to the rest of the project.
This is an indication that warning density related metrics can be of use in just-in-time defect prediction scenarios, i.e., change based scenarios, as also shown by~\cite{Querel2021} and in our previous work~\cite{Trautsch2020a}.
However, the effect is negligible in our data. This was also the case for predictive models by~\cite{Querel2021}.
Thus, any gain in prediction models due to general static analysis warnings is likely very small.

However, when we look at the median difference between bug inducing files and the rest of the project at that point in time we see that bug inducing files contain less static analysis warnings.
This counter intuitive result can be fully explained by the overall decreasing warning density over time we found in our previous study~\citep{Trautsch2020}.
This finding is highly relevant for researchers, because this shows the importance of accounting for time as confounding factor for the evaluation of the effectiveness of methods.
Without the careful consideration of the change over time, we would now try to explain why bug inducing files have less warnings and other researchers may built on this faulty conclusion.
Therefore, this part of our results should also be a cautionary tale for other researchers that investigate the effectiveness of tools and methods: if the complete project is used as a baseline, it should always be considered when source code was actually worked on.
If parts of the source code have been stable for a long time, they are not suitable for a comparison with recently changed code, without accounting for general changes, e.g., in coding or testing practices, over time.

However, we did find that code with more PMD warnings leads to more bugs when changed.
When looking into the differences between bug inducing file changes and all other file changes we find significant differences in 3 of 4 cases.
While the effect size is negligible, in all cases using only the default rules yields a higher effect size.
These rules were hand-picked by the Maven developers, arguably because of their importance for the internal quality.
For practitioners, this finding if of particular importance: not only does it reduce the number of alerts to carefully select ASAT warnings from a large set of candidates, it also helps to reduce general issues that are associated with bugs.

This also has implications for researchers when including warning density based metrics into predictive models. Our data shows that the model might be improved by choosing an appropriate subset of the possible warnings of an \ac{ASAT}.
Using all warnings without considering their potential relation to defects is not a good strategy.
Our data also shows that a good starting point might be a commonly used default, e.g., for PMD the maven-pmd-plugin default rules.

\section{Threats to validity}\label{sec:threats_to_validity}
In this section, we discuss the threats to validity we identified for our work.
To structure this section we discuss four basic types of validity separately, as suggested by~\cite{wohlin}.

\subsection{Construct validity}
A threat to the relation between theory and observation may occur in our study from the measurement of warning density.
We restrict the data to production code to mitigate effects test code has on warning density as it is often much simpler than production code.

\subsection{Internal validity}
A general threat to internal validity would be a selection of static analysis warnings.
We mitigate this by measuring the warning density for all warnings and for only default warnings as a common subset for Java projects.
Due to the nature of our approach, we mitigate differences between projects regarding the handling of warnings as well as the impact of size.

\subsection{External validity}
Due to the usage of manually validated data in our study, our study subjects are restricted to those for which we have this kind of data.
This is a threat to the generalizability of our findings, e.g., to all Java projects or to all open source projects.
Still, as we argue in \citep{line_label}, our data should be representative for mature Java open source projects.

Moreover, we observe only one static analysis tool (PMD). While this may also restrict the generalizability of our study, we believe that due to the large range of rules of this \ac{ASAT} our results should generalize to \acp{ASAT} that are broad in scope. \acp{ASAT} of a different focus, e.g., on coding style (Checkstyle) of directly finding bugs (FindBugs, SpotBugs) may result in different results.

\subsection{Conclusion validity}
We report lower warning density for bug inducing files in comparison to the rest of the project at that point in time.
While this reflects the difference in warning density between the file and the project, it can be influenced by constantly decreasing warning density.
We mitigate this by also including a comparison between bug inducing changes and all other changes.

\section{Conclusion}\label{sec:conclusion}
In this article we provide evidence for a common assumption in software engineering, i.e., that static analysis tools provide a net-benefit to software quality even though they suffer from problems with false positives.
We use an improved state-of-the art approach used for fine-grained just-in-time defect prediction to establish a link between files within commits that induce bugs and measure warning density related features which we aggregate over the evolution of our study subjects.
This approach runs on data which allows us to remove several noise factors from our data, wrong issue types, wrong issue links to commits and tangled bug fixes.
The analysis approach allows us to merge the available data as it mitigates differences between projects, sizes and to some extend the evolution of warnings over time.

We find that bugs are induced in files which have a comparably low warning density, i.e., less static analysis warnings than the files of the rest of the project at the time the bug was induced.
However, this difference can be explained by the fact that the warning density decreases over time.
When we compare the bug inducing changes with all other changes, we do find a significant higher warning density when using all PMD rules in one of two metrics.
However, the effect size is negligible.
When we use a small rule set that restricts the \numberPMDWarnings{} PMD warnings to the \numberDefaultWarnings{} warnings hand-picked by the Maven developers as default warnings, we find that bug inducing changes have a significant but also negligible larger warning density.
However, the effect size increases for the default rule set.
Assuming that the smaller rule set was crafted with the intent to single out the most important rules for the quality, this indicates that there is indeed a (weak) relationship between general ASAT tools and bugs.

This is also direct evidence for a common best practice in the use of static analysis tools:
Appropriate rules for \acp{ASAT} should be chosen for the project.
This not only reduces the number of alarms, which is important for the acceptance by developers, but also has a better relationship with the external quality of the software measured through bugs.

\section*{Declarations}
This work was partly funded by the German Research Foundation (DFG) through the project DEFECTS, grant 402774445.\\
The authors have no competing interests to declare that are relevant to the content of this article.

\bibliographystyle{spbasic}      
\bibliography{literature}

\end{document}